\def\parno {\par\noindent}
\magnification=\magstep1
\tolerance=20000

\centerline {\bf {TRANSFER MATRICES, NON-HERMITIAN HAMILTONIANS }}
\centerline {\bf {AND RESOLVENTS: SOME SPECTRAL IDENTITIES}}
\vskip 0.5truecm
\centerline {Luca Molinari}
\centerline {Dipartimento di Fisica dell'Universit\`a di Milano and I.N.F.N.}
\centerline {Via Celoria 16, 20133 Milano}
\vskip 0.5truecm
{\bf Abstract.} I consider the $N$-step transfer matrix $T$ for a general block 
Hamiltonian, with eigenvalue equation 
$$ L_n\psi_{n+1}+H_n\psi_n + L_{n-1}^\dagger\psi_{n-1}=E\psi_n $$ 
where $H_n$ and $L_n$ are matrices, and provide its explicit representation
in terms of blocks of the resolvent of the Hamiltonian matrix for the system
of length $N$ with boundary conditions $\psi_0=\psi_{N+1}=0$. I then introduce
the related Hamiltonian for the case $\psi_0=z^{-1}\psi_N$ and $\psi_{N+1}=
z\psi_1$, and provide an exact relation between the trace of its 
resolvent and Tr$(T-z)^{-1}$, together with an identity of Thouless type 
connecting Tr $(\log |T|)$ with the Hamiltonian eigenvalues for $z=e^{i\phi}$. 
The results are then extended to $T^\dagger T$ by showing that it is itself a 
transfer matrix. Besides their own mathematical interest, the identities should 
be useful for an analytical approach in the study of spectral properties of a 
physically relevant class of transfer matrices.\parno
P.A.C.S.: 02.10.Sp (theory of matrices), 05.60 (theory of quantum transport),
71.23 (Anderson model), 72.17.Rn (Quantum localization)
\vskip 0.5truecm\parno
To appear in October 1998, on J. Phys. A: Math. Gen.  
\vskip 5truecm\noindent
E-Mail address: luca.molinari@milano.infn.it
\vfill\eject

{\bf {\S1 Introduction.}}\par 
Several discretized models are described by a Hamiltonian matrix ${\cal H}$ 
with tridiagonal structure made of blocks $H_n=H_n^\dagger $ along the 
main diagonal, and blocks $L_n$, $L_n^\dagger $, with $\det L_n\neq 0$,
respectively in the adjacent upper and lower diagonals, the blocks having size 
$M\times M$. The diagonal matrices may describe the inner dynamics of a 
sequence of finite subsystems, and the off-diagonal matrices are the couplings
among neighbouring ones.\parno
An important and extensively studied example is Anderson's model for 
electronic transport in a D-dimensional lattice with random potential, which 
for D=3 exhibits a metal-insulator transition [1]. The single matrices $H_n$ 
are random Hamiltonians for the isolated slices of dimension D-1 and, in the 
simplest case, the couplings $L_n$ are proportional to the unit matrix, as 
implied by the discretization of the Laplacian. They may also be complex, in 
the presence of a magnetic field [2], or random, due to random hopping 
amplitudes [3]. Another frequently studied model is the ensemble of band 
random matrices [4], where $H_n$ is a member of GOE or GUE and $L_n$ are 
random and lower triangular matrices. They found applications in quantum 
chaos [5], one particle mesoscopic transport [6] and the propagation of two
particles in disordered media [7]. The block structure also arises in 
the Fourier representation of the Floquet Hamiltonian $H_t=H_0+ V^\dagger 
e^{i\omega t} + V e^{-i\omega t}-i\partial_t $, giving $H_n=H_0+n\omega $ and 
$L=V$. Block Hamiltonians have also been investigated in the context of matrix 
models [8], the blocks being rotationally invariant for the methods to apply.
\par
The eigenvalue equation for ${\cal H}$, in block components, is:
$$ L_n \psi_{n+1} + H_n\psi_n +  L_{n-1}^\dagger \psi_{n-1} = E\psi_n 
\eqno {(1.1)}$$
One is often interested in asymptotic properties of eigenvectors. A
basic tool for this analysis, which exploits the recursive content of the 
eigenvalue equation, is the transfer matrix, connecting pairs of components 
of the vector 
$$ \pmatrix {L_N\psi_{N+1}\cr \psi_N\cr } = T(E)
   \pmatrix {\psi_1\cr L_0^\dagger\psi_0\cr } \eqno {(1.2)} $$ 
The transfer matrix has size $2M\times 2M$ and it is convenient to
factorize it as follows:
$$ T(E)= T_N(E) \Sigma_{N-1} T_{N-1} (E)\Sigma_{N-2}\ldots \Sigma_1 T_1(E) 
\eqno {(1.3)} $$
$$ T_k(E) = \pmatrix {E-H_k & -I\cr I & 0\cr }, \quad \Sigma_k= \pmatrix {
L_k^{-1} & 0 \cr 0 & L_k^\dagger \cr }  $$
By allowing for complex values of the parameter $E$, for the single factors
and then for the whole product, one obtains the important symplectic property
$$ T(E^*)^\dagger \sigma_2 T(E) = \sigma_2 , \quad \quad 
\sigma_2 =\pmatrix {0 & -I \cr I & 0\cr } \eqno {(1.4)} $$
where $I$ is the unit matrix of size $M$. Another consequence of the factorized
expression is $\det T(E) = \prod_k \det [L_k^\dagger L_k^{-1}] $, that 
implies $|\det T(E)|=1$.\par
General theorems assert that the eigenvalues $t_a$ of transfer matrices built 
with random factors grow or decrease exponentially with $N$ [9], allowing
the definition of characteristic exponents
$$ \gamma_a(E) = \lim_{N\to\infty }{1\over {N}}\log |t_a(E)| \eqno {(1.5)} 
$$
which, in the case of $(T^\dagger T)^{1/2}$, constitute the Lyapunov spectrum
of the model. The analytical derivation of a Lyapunov spectrum is usually
extremely difficult, the alternative being a careful numerical work to cope 
with exponential instabilities. For $2\times 2$ matrices ($M=1$), a relevant 
formula by Herbert and Jones, rediscussed by Thouless and bearing his name, 
connects the single Lyapunov exponent to the eigenvalue density of the ensemble 
of Jacobi Hamiltonians [10]. However, the density is by no means a simpler
problem; an exception is Lloyd's model, characterized by diagonal 
disorder with Cauchy distribution: in this case the analytical expression
of $\gamma (E)$ is known [10]. The statistical properties of
the Lyapunov exponent and various generalizations have been investigated
extensively by Pendry [11].\parno
One of the rare solvable examples in more than one dimension is in the work 
[12] by Isopi and Newman, who studied products of matrices all of whose 
entries are identically distributed random variables, and found analytically a 
"triangle law" for the Lyapunov spectrum; Cook and Derrida considered 
the case of randomly sparse matrices [13]. A beautiful statistical theory of 
transfer matrices, based on few physical contraints, has been introduced by 
Mello and others to describe transport properties in disordered multichannel
conductors , obtaining the observed value for universal conductance 
fluctuations [14]. In general, however, the transfer matrix is a derivated
object, which inherits a structure from the specific recurrence or dynamical
equation under examination. Its statistical properties depend in a 
complicated way on fluctuating parameters that enter more naturally, for
example, in the characterization of an ensemble of Hamiltonians. \parno
For the class of Hamiltonians we are considering, analytical results are 
lacking. Many extensive numerical calculations have been carried out for the 
Anderson model. It was by means of transfer matrices that  Kramer and 
MacKinnon first exhibited the metal-insulator transition in 3 dimensions [15], 
through the different scaling behaviour in the transverse area $M$
of the smallest Lyapunov exponent. A recent numerical study by Markos 
[16] provides the whole Lyapunov spectrum, which is sensitive to the 
transition. The Lyapunov spectrum of band random matrices was obtained 
numerically by Kottos et al. [17], with a discussion of finite size 
corrections.\par
The aim of this paper is to investigate some general mathematical 
properties of the transfer matrix $T(E)$ that arise from the block structure 
of a single but generic Hamiltonian matrix ${\cal H}$, of size $NM$, 
corresponding to (1.1) with boundary condition $\psi_0=\psi_{N+1} =0$.\parno
In [18] I showed that the eigenvalues of the transfer matrix 
are most directly related to those of a matrix ${\cal H}(z)$ of size $NM$, 
which in general is not Hermitian and has block structure
$$ {\cal H}(z) = \pmatrix {
H_1            & L_1 & {} & {} & {} & {} & {} & {{1\over z}I} \cr
L_1^\dagger   & H_2 & L_2 & {} & {} & {} & {} & {} \cr
{}    & L_2^\dagger & \ldots & {} & {} & {} & {} & {} \cr
{}    &    {}       &        & {} & {} &\ldots  & H_{N-1} & L_{N-1} \cr
{zI}  &    {}       &        & {} & {} & {} & L_{N-1}^\dagger & H_N \cr }
\eqno {(1.6)}  $$
resulting from the eigenvalue equation (1.1) with boundary conditions specified
through a complex parameter $z$: 
$$  L_N\psi_{N+1}= z\psi_1  \quad L_0^\dagger\psi_0=
{1\over z}\psi_N   \eqno {(1.7)} $$
The relation is based on the simple statement, whose proof is straightforward:
\parno
{\sl A vector $(\psi_1,\psi_2, \ldots,\psi_N)$  is an eigenvector 
of ${\cal H}(z)$ with eigenvalue $E$ if and only if $(z\psi_1, \psi_N)$ is an 
eigenvector of $T(E)$ with eigenvalue $z$, the components $\psi_2\ldots 
\psi_{N-1}$ being linked to $\psi_N$ and $L_N\psi_{N+1}=z\psi_1$ by (1.1).}
\parno
It implies that the characteristic polynomials of $T(E)$ and 
${\cal H}(z)$ are proportional, and eventually leads to the following 
"duality relation":
$$ \det [T(E)-z] = (-z)^M \det(L_{N-1}\ldots L_1)^{-1}\det [E-{\cal H}(z)]
\eqno {(1.8)} $$
A new proof will be given in \S2, after having derived  an explicit 
representation of the matrix $T(E)$ in terms of the corner blocks $G_{ij}$, 
$i,j=1,N$, of the resolvent $G=({\cal H}-E)^{-1}$ of the Hamiltonian matrix
${\cal H}$, for the system of length $N$. The Hamiltonian has block structure 
(1.6) with null matrices replacing the $z$-dependent corner blocks, 
corresponding to the boundary requirement $\psi_0=\psi_{N+1}=0$.\parno
In this paper it is noted that the derivative in the variable $z$ gives a 
relation between resolvents
$$ {\rm Tr} \left ( {1\over {T(E)-z}} \right ) = - {M\over z} + 
{\partial\over {\partial z}}\log \det [E-{\cal H}(z)] \eqno {(1.9)} $$
The equation also follows from a representation of $[T(E)-z]^{-1}$ in terms
of the corner blocks of the resolvent $\tilde G = ({\cal H}(z)-E)^{-1}$, 
to be obtained in \S4. Did $\cal H$ belong to an ensemble, it would provide 
access to the spectral density of $T(E)$ by relating the difficult problem of 
averaging the resolvent of the transfer matrix, which depends on the ensemble 
parameters in a complex way, to the average of a ratio of determinants of the 
Hamiltonian itself. \parno
The general discussion for $T(E)$ will be extended in \S5 to cover the 
relevant matrix $T(E)^\dagger T(E)$, by showing that it is itself the 
transfer matrix of a tridiagonal block Hamiltonian, of size $2NM$.\parno
The main results of the paper can be summarized in the following 
equations: relation (1.9) or its equivalent (3.6) among traces of resolvents, 
(2.3) and (4.3) that provide the representation of the transfer matrix in 
terms of corner blocks of resolvents, (3.5) and (5.4) that generalize 
Thouless' relation to $M>1$ and relate Lyapunov exponents to the eigenvalues 
of the Hamiltonian matrix, the duality relation (5.6) for ther matrix 
$T^\dagger T$.\par
The described results are exact and general. Hopefully, this work should 
provide an analytical framework for the hard task of investigating average 
spectral properties of transfer matrices, constructed  from an ensemble of 
Hamiltonians of this type. One more motivation  is the recent interest in 
Non-Hermitian matrices, which are now extensively investigated mainly 
in the one-dimensional case $M=1$ [19], precisely with the tridiagonal 
structure and boundary conditions that are here generalized. In one dimension,
the eigenvalues are distributed along curves in the complex plane [20]
and real eigenvalues correspond to delocalized eigenvectors [21]. The here
established relationship with the transfer matrix makes non Hermitian 
matrices an obvious object for investigating localization properties of 
eigenvectors. For $M=1$ this was done by Brouwer et al. [22]. \par
\par\vskip 0.5truecm

{\bf {\S2 The transfer matrix and the resolvent of ${\cal H}$. }}\par
In this section I obtain a block representation of $T(E)$ in terms of the 
corner blocks of the resolvent of ${\cal H}$. I then prove the duality 
relation (1.8).\parno 
For complex $E$, let us introduce the resolvent $G(E)=({\cal H}-E)^{-1}$. It
is a matrix made of $N^2$  square blocks $G_{ij}$ of size $M\times M$. 
The relation $[G(E)_{ij}]^\dagger = G(E^*)_{ji}$ holds. By definition:
$$ L_{i-1}^\dagger G_{i-1,j} + (H_i-E)G_{i,j} +L_i G_{i+1,j} = I\delta_{ij} 
\eqno {(2.1)} $$
By solving the recurrence relations for $j=1$ and $j=N$, one obtains two
identities involving the transfer matrix
$$ \pmatrix {0\cr G_{N,1}\cr }=T(E) \pmatrix {G_{1,1}\cr -I\cr },\quad 
\pmatrix {-I\cr G_{N,N}\cr }=T(E) \pmatrix {G_{1,N}\cr 0\cr } \eqno {(2.2)} $$
They can be joined into a matrix relation, which gives a representation of the 
transfer matrix in terms of the corner blocks of the resolvent:
$$ \eqalignno {T(E) = & \pmatrix {-I & 0\cr G_{N,N} & G_{N,1}\cr }
\pmatrix {G_{1,N} & G_{1,1}\cr 0 & -I\cr }^{-1}=  & {(2.3a)} \cr
=& \pmatrix {-G_{1,N}^{-1} & -G_{1,N}^{-1}G_{1,1}\cr G_{N,N}G_{1,N}^{-1} & 
-G_{N,1}+ G_{N,N}G_{1,N}^{-1}G_{1,1} } & (2.3b) \cr }$$
One checks that the symplectic property (1.4) is identically satisfied.
Note that each block component of $T(E)$ is a matrix polynomial in $E$, and is
here explicitly represented in terms of the resolvent of the Hamiltonian. 
By varying the number $N$ of factors in $T(E)$, one obtains a collection of
matrix polynomials which follow orthogonality relations that extend to $M>1$ 
the familiar notion of othogonal polynomials of Jacobi matrices [23].\par
To the end of deriving the duality relation, I first prove 
$$ \det G_{1,N}^{-1} = \det [L_1\ldots L_{N-1}]^{-1} \det [E-{\cal H}]
\eqno {(2.4)} $$
Proof: a vector $(\psi_1, \ldots ,\psi_N)$ is an eigenvector of ${\cal H}$ 
with eigenvalue $E$ if and only if it solves (1.1) with boundary conditions 
$\psi_{N+1}=\psi_0=0$. These conditions, by eqs. (1.2) and (2.3b), are 
equivalent to the requirement $0=G_{1,N}^{-1}\psi_1 $. By construction $T(E)$
is a polynomial in $E$ of degree $N$ with matrix coefficients; then
$\det G_{1,N}^{-1}$ is a polynomial in $E$ of degree $NM$ and leading term
$(-1)^ME^{MN}\det[L_{N-1}^{-1}\ldots L_1^{-1}]$. The polynomials $\det [E-
{\cal H}]$  and $\det [G_{1,N}^{-1}]$ share all zeros and are therefore
proportional by a numerical factor obtained from comparison of the leading 
terms.$\bullet $ \parno
I now give a proof of the duality relation (1.8), alternative to that
provided in [18].\parno
Proof: by writing ${\cal H}(z)={\cal H}+V(z)$, where $V(z)$ is zero 
everywhere except for the corner blocks $V_{1,N}={1\over z}I$ and $V_{N,1}=zI$,
one calculates:
$$ \eqalign {\det [E-{\cal H}(z)]=&\det [E-{\cal H}] \det [I+G(E)V(z)] =\cr
=& \det [E-{\cal H}]\det \pmatrix {I+zG_{1,N} & {1\over z}G_{1,1}\cr
zG_{N,N} & I+{1\over z}G_{N,1} \cr } \cr } \eqno {(2.5)} $$
On the other hand, by using algebraic properties of determinants, one obtains
from eq. (2.3b):
$$ \det [T(E)-z] = z^M \det [G_{1,N}^{-1}] \det \pmatrix
{I+zG_{1,N} & {1\over z} G_{1,1}\cr zG_{N,N} & I+{1\over z}G_{N,1} \cr }
\eqno {(2.6)}  $$
By taking into account property (2.4), the duality relation (1.8) follows.
$\bullet $\parno
\vskip 0.5truecm

{\bf {\S3 The duality relations.}}\par
I here discuss some consequences of the duality relation, 
$$ \det [T(E)-z] = (-z)^M \det(L_{N-1}\ldots L_1)^{-1}\det [E-{\cal H}(z)]
\eqno {(3.1)} $$
An identity for the inverse transfer matrix is obtained from the adjoint of 
(3.1), with the aid of the properties $T(E^*)^\dagger =-\sigma_2 T(E)^{-1}
\sigma_2 $ and ${\cal H}(z^*)^\dagger = {\cal H}(1/z)$:
$$ \det [T(E)^{-1}-z] = (-z)^M \det(L_{N-1}^\dagger \ldots L_1^\dagger )^{-1}
\det [E-{\cal H}(1/z)]. \eqno {(3.2)} $$
The product of the two identities immediately yields an identity which was 
used in  [18] to study the structure of bands and the dynamics of eigenvalues
of the Hermitian Hamiltonians ${\cal H}(e^{i\varphi})$:
$$ \det [T(E)+T(E)^{-1}-(z+{1\over z})] = |\det(L_{N-1} \ldots L_1 )|^{-2}
\det [E-{\cal H}(z)]\det [E-{\cal H}(1/z)] \eqno {(3.3)} $$
A simple general consequence of (3.1) is that, for Im$E\neq 0$, the transfer 
matrix $T(E)$ has no eigenvalues on the unit circle, since the right term of 
(3.1) never vanishes for a Hermitian matrix. More generally, this is true for 
$E$ not in the union of the bands $B_k$, $k=1\ldots NM$, each one being
defined as the interval of the real axis spanned by the eigenvalue 
$E_k(\varphi )$ of the Hermitian matrix ${\cal H}(e^{i\phi })$, as $\varphi $ 
varies in $[0,2\pi)$.\parno
From the symplectic property (1.4) it follows that if $t$ is an eigenvalue of 
$T(E)$ with $|t|\neq 1$, then $1/t^*$ is an eigenvalue of $T(E^*)$. In 
particular, for real $E$, the $2M$ eigenvalues of $T(E)$ occur in pairs $t$, 
$1/t^*$, unless $|t|=1$. The number $\nu $ of pairs of eigenvalues on the 
unit circle
coincides with the number of bands with intersection in $E$ [18].\parno
Let us denote the eigenvalues of $T(E)$, $E$ real, as 
$$   t_a= e^{\pm \lambda_a +i\theta_a}, \quad a=1,\ldots ,M-\nu,  \quad\quad
t_b=e^{i\theta_b}, \quad b=1,\ldots ,2\nu $$
and write the modulus of (3.1), with real $E$ and $z=e^{i\varphi }$, in terms 
of them. After some simple algebra:
$$ \prod_{a=1}^{M-\nu} \left ( 
2\cosh \lambda_a -2\cos (\theta_a -\varphi )\right )
\prod_{b=1}^{2\nu } 2 \sin \left ({1\over 2}|\theta_b-\varphi |\right ) =$$
$$ = |\det(L_{N-1}\ldots L_1 )|^{-1} |\det [E-{\cal H}(e^{i\varphi })] |
\eqno {(3.4)} $$
By taking the logarithm of it and integrating the phase
$\varphi $ in $[0,2\pi )$ one obtains a remarkably simple and interesting 
relation:
$$ \sum_{a=1}^{M-\nu} \lambda_a (E)= -\sum_{j=1}^{N-1}\log |\det L_j| +
{1\over {2\pi}} \int_0^{2\pi} d\varphi \log |\det [E-{\cal H}(e^{i\varphi})]|
\eqno {(3.5)} $$
This formula is exact, and is valid for a single matrix. In the large $N$ 
limit and in a statistical context, one would have the
average behaviour $\langle \lambda_a\rangle =N\gamma_a$, where $\gamma_a $ is 
independent of $N$. The right hand side of (3.5) would be evaluated
by means of the ensemble and $\varphi $-averaged density of eigenvalues of
${\cal H}(e^{i\varphi })$. The equation would then provide a generalization
to $M>1$ of the Thouless relation between the Lyapunov spectrum and the  
average eigenvalue density of the Hamiltonian ensemble [10]. \par
If instead we take in (3.1) the derivative in the variable $z$, by using
the property $d/dz \det (A+zI)=\det (A+zI) {\rm tr}(A+zI)^{-1}$, we obtain 
equation (1.9). Also the derivative in the right hand side can be 
computed, and gives the following final formula, where ${\tilde G}_{ij}$ are 
the blocks of size $M\times M$ that partition the resolvent $\tilde G (z,E)=
[{\cal H}(z)-E]^{-1} $:
$$ {\rm Tr} \left ( {1\over {T(E)-z}} \right ) = - {M\over z} -
{\rm Tr} \tilde G_{1,N} + {1\over z^2} {\rm Tr}\tilde G_{N,1} \eqno {(3.6)} $$
The same formula follows from a stronger result, to be given in the next
section.
\vskip 0.5truecm

{\bf {\S4 The transfer matrix and the resolvent of ${\cal H}(z)$. }}\par
A representation of the resolvent of the transfer matrix 
can be given in terms of the resolvent of the non-Hermitian matrix, 
$ \tilde G(z,E)=[{\cal H}(z)-E]^{-1}$. Note that $[{\tilde G}(z,E)_{ij}]
^\dagger = \tilde G(1/z^*,E^*)_{ji}$. With the same procedure as in \S2, one 
obtains two identities
$$ \pmatrix { z{\tilde G}_{1,1}\cr {\tilde G}_{N,1} \cr }= T(E)
\pmatrix { {\tilde G}_{1,1} \cr {1\over z}{\tilde G}_{N,1}-I \cr },\quad
\pmatrix {z{\tilde G}_{1,N} -I\cr {\tilde G}_{N,N}\cr }= T(E)
\pmatrix {{\tilde G}_{1,N}\cr {1\over z}{\tilde G}_{N,N} \cr } \eqno {(4.1)}$$
which join into the matrix relation
$$ \pmatrix {z{\tilde G}_{1,N}-I & z{\tilde G}_{1,1}\cr {\tilde G}_{N,N} &
{\tilde G}_{N,1} \cr } = T(E) \pmatrix { {\tilde G}_{1,N} & {\tilde G}_{1,1}\cr
{1\over z} {\tilde G}_{N,N} & {1\over z} {\tilde G}_{N,1}-I \cr } 
\eqno {(4.2)}$$
Simple steps lead to the final representation:
$$ {1\over {T(E)-z}} 
= \pmatrix { -{\tilde G}_{1,N} & {1\over z} {\tilde G}_{1,1}
\cr -{1\over z}{\tilde G}_{N,N} & {1\over {z^2}} {\tilde G}_{N,1}-{1\over z}I
\cr } \eqno {(4.3)} $$
which, by taking the trace, provides eq. (3.6). \parno
Note that the corner blocks of ${\tilde G}$ can be expressed in closed form in 
terms of the corner blocks of $G$, by means of the Lippman Schwinger equations 
$$ G_{i,j} ={\tilde G}_{i,j} + {1\over z}G_{i,1}{\tilde G}_{N,j} +
z G_{i,N}{\tilde G}_{1,j}\eqno {(4.4)} $$
\vskip 0.5truecm

{\bf {\S5 The matrix $T^\dagger T$.}}\par
The general results obtained so far for transfer matrices can also be applied 
to the matrix $Q(E)\equiv T(E)^\dagger T(E)$, which will be shown to be itself
the transfer matrix of a Hamiltonian built out of ${\cal H}$.\parno
The matrix $Q(E)$ has the feature of being Hermitian and positive, therefore 
with real and positive eigenvalues. It is easy to show the validity of 
the property:
$$        Q(E^*)\sigma_2 Q(E) =\sigma_2 \eqno {(5.1)} $$
It follows that if $t$ is an eigenvalue of $Q(E)$, then $1/t$ is eigenvalue of
$Q(E^*)$.\parno
While considering the factorization
$$ T(E)^\dagger T(E)= T_1(E)^\dagger \Sigma_1^\dagger T_2(E)^\dagger\ldots  
\Sigma_{N-1}^\dagger T_N(E)^\dagger T_N (E) \Sigma_{N-1} \ldots \Sigma_1
 T_1(E) $$ 
one notes the property that $T_k(E)^\dagger $, constructed with $H_k$, 
coincides with $-T_k(-E^*)$ constructed with $-H_k$. This allows to interpret
$Q(E)$ as the transfer matrix for the solution of the equation 
${\cal K}(E)\Psi =0$, with matrix ${\cal K}(E)=$
$$ = \pmatrix { 
H_1-E       & L_1         & {}      & {}     & {} & {} & {}\cr
L_1^\dagger & H_2-E       & {}      & {}     & {} & {} & {}\cr
{}          & L_2^\dagger & \ldots  & {} & {}& {} & {} & {}\cr
{}          & {}          &  {}              & L_{N-1}& {} & {} & {}\cr
{}          & {}          &  L_{N-1}^\dagger & H_N-E  & -I  & {} & {}\cr
{}          & {}          &  {}              & -I & E^*-H_N & -L_{N-1}^\dagger
& {}\cr
{}          & {}          & {}               &{} & -L_{N-1} & E^*-H_{N-1}
& -L_{N-2}^\dagger & {}\cr
{}          & {}          & {}               &{} &      {}  & -L_{N-2} 
&    {}            & {}\cr
{}          & {}          & {}               &{} &      {}  &   {}
& \ldots           & -L_1^\dagger\cr
{}          & {}          & {}               &{} &      {}  &   {}
& -L_1             & E^*-H_1 \cr
} $$
The corresponding non-Hermitian matrix ${\cal K}(E,z)$ entering the 
duality relation, is obtained by placing the diagonal matrices $z^{-1}I$ and 
$zI$ in the upper right and lower left corners respectively. Then, the 
following equation holds:
$$ \det [T(E)^\dagger T(E) -z] = (-1)^{NM}(-z)^M |\det(L_{N-1}\ldots L_1)|^{-2}
\det {\cal K}(E, z) \eqno {(5.2)} $$
A few remarks on the spectral properties of ${\cal K}(E,z)$, which can
be easily translated for the matrix ${\cal K}(E)$, are useful:
\parno
\item {a)}{ ${\cal K}(E,z)^\dagger = {\cal K}(E^*, 1/z^*) $;}
\item {b)}{ $\pmatrix {0 & P\cr -P & 0\cr }{\cal K}(E,z)\pmatrix {0 & -P\cr P 
  & 0\cr }= -{\cal K} (E^*, 1/z) $\par
 where $P$ is the block matrix with nonzero blocks $P_{i,N-i+1}=I$ of 
 size $M$, $i=1\ldots N$;}
\item {c)}{for real $z$, because of a) and b), the eigenvalues of 
${\cal K}(E,z)$ come in pairs $x $, $-x^*$;}
\item {d)}{the value $x=0$ does not belong to the spectrum of 
${\cal K}(E,z)$ if $z$ is not in the real positive axis, because the left side
in (5.2) cannot vanish.}\parno
Let us concentrate on the case where $z=e^{i\varphi }$ and $E$ is real;
the matrix ${\cal K}(E,e^{i\varphi })$ is Hermitian, therefore its eigenvalues
are real. By point c) the eigenvalues for $\varphi =0,\pi $ are symmetric; they
also mark the extrema of the bands [18]: it follows that 
${\cal K}(E,e^{i\varphi})$ has as many positive as many negative eigenvalues
and $(-1)^{NM}\det  {\cal K}(E,e^{i \varphi }) >0$. 
The eigenvalues of $Q(E)$ are $M-\mu$ positive pairs $(t_a,t_a^{-1})$, with
$t_a=e^{\lambda_a}>1$, being $2\mu $ eigenvalues equal to unity. Equation 
(5.2) reads, in terms of the eigenvalues of $Q(E)$:
$$ \left ( 2\sin {\varphi\over 2} \right )^{2\mu } \prod_{a=1}^{M-\mu }
\left (2\cosh \lambda_a -2\cos\varphi \right ) = 
\prod_{k=1}^{N-1} |\det L_k|^{-2}(-1)^{NM} \det {\cal K}(E,e^{i\varphi }) 
\eqno {(5.3)} $$
By taking the logarithm and integrating in $\varphi $, we end with a formula
of Thouless type:
$$ \sum_a \lambda_a(E) = -2\sum_{k=1}^{N-1} \log |\det L_k | +
{1\over {2\pi }}\int_0^{2\pi } d\varphi \log [(-1)^{NM}\det {\cal K} (E,
e^{i\varphi }) ] \eqno {(5.4)} $$
For real $z$ and $E$ real or complex it is convenient, at 
least for a simplification of the notation, to bring the 
matrix ${\cal K}(E,z)$ to another form; there is much freedom since only
the determinant of the matrix matters. Let us choose to left and right-multiply
matrix ${\cal K}(E,z)$ by unitary matrices to give:
$$ \eqalign { {\cal K}^\prime (E,z)\equiv & {1\over {\sqrt 2}}
\pmatrix {I & P\cr -iI & iP \cr }{\cal K}(E,z){1\over {\sqrt 2}}
\pmatrix {-I & iI\cr P & iP \cr }\cr
= & \pmatrix { {\cal H} -{\rm Re}E + U & -iV-{\rm Im}E \cr iV -{\rm Im}E 
& -{\cal H}+ {\rm Re}E+ U \cr }  \cr } \eqno {(5.5)} $$
where $P$ is the same matrix of size $NM$ defined after (5.2), $U$ and
$V$ are block diagonal matrices, each of the $N$ diagonal blocks having size 
$M$. The only nonzero blocks are: $U_{1,1}= {1\over 2}(z-z^{-1})I$, $V_{1,1}=
{1\over 2}(z+z^{-1})I$ and $V_{N,N}= -I$. With this transformation, we 
obtain the equivalent form of the duality relation:
$$ \det [T(E)^\dagger T(E) -z] = (-1)^{NM}(-z)^M |\det(L_{N-1}\ldots L_1)|^{-2}
\det {\cal K}^\prime (E, z) \eqno {(5.6)} $$
The matrix ${\cal K}^\prime (E,z)$ is Hermitian for any $E$ in the complex 
plane and  real $z$, and it has the advantage of containing the matrix 
${\cal H}$ in the diagonal blocks, albeit with opposite sign; a similar 
structure appears in a paper by Efetov [24]. It has the following properties:
\item {a)}{ ${\cal K}^\prime (E,z)^\dagger = {\cal K}^\prime (E,z^*)$ }
\item {b)}{ $\pmatrix {0& I\cr I & 0\cr } {\cal K}^\prime (E,z) 
\pmatrix {0& I\cr I & 0\cr }= -{\cal K}^\prime (E^*, 1/z) $ }
\item {c)}{ $\pmatrix {I& 0\cr 0 & -I\cr } {\cal K}^\prime (E,z) 
\pmatrix {I & 0\cr 0 & -I\cr }= {\cal K}^\prime (E^*, -1/z) $ }\par
\vskip 0.5truecm

{\bf {\S6 Conclusions.}}\par
In the present paper and in [18] I have considered a class of Hamiltonians
characterized by a block Jacobi structure which is shared in many interesting 
models of quantum disordered transport. For a single Hamiltonian matrix I
have obtained exact relations that allow to describe spectral properties
of transfer matrices through properties of the Hamiltonian itself. The 
identities involve general boundary conditions that imply a close connection 
between transfer matrices and Non-Hermitian Hamiltonians, and make the 
statistical analysis less involved, since the statistical ensemble is usually 
defined for the Hamiltonian.\parno
Some equations, like (1.9), are suited for the supersymmetric technique. Two 
problems arise, that have already been considered in the literature [25]: 
1) the need of a special formalism for the determination of the density of 
complex eigenvalues from the knowledge of the average resolvent, 2) the 
"Hermitianization" procedure for representing ratio of determinants, which
provide by differentiation traces of resolvents, as  Gaussian superintegrals.  
These problems are absent while considering the relation for $T^\dagger T$.\par
\vfill\eject

{\bf {References.}}\parno
\item {[1]}{B.Kramer and A.MacKinnon: "Localization: theory and 
experiment", Rep. Progr. Phys. 56 (1993) 1469.}
\item {[2]}{T.Dr\"ose, M.Batsch, I.Zharekeshev and B.Kramer: "Phase diagram
of localization in a magnetic field", Phys. Rev. B 57 (1998) 37.}
\item {[3]}{A.Eilmes, R.A.R\"omer and M.Schreiber: "The two-dimensional
Anderson model of localization with random hopping" Eur. Phys. J. B 1 (1998)
29;\par
T.Kawarabayashi, B.Kramer and T.Ohtsuki: "Anderson transition in 
three-dimensional disordered systems with randomly varying magnetic flux",
Phys. Rev. B 57 (1998) 11842.}
\item {[4]}{Y.V.Fyodorov and A.D.Mirlin: "Scaling properties of localization
in Random Band Matrices: a $\sigma-$model approach", Phys. Rev. Lett. 67 
(1991) 2405;\par
P.G.Silvestrov: "Summing graphs for random band matrices", Phys. Rev. E 55
(1997) 6419.}
\item {[5]}{G.Casati, B.V.Chirikov, I.Guarneri and F.M.Izrailev: "Quantum
ergodicity and localization in conservative systems: the Wigner band random 
matrix model", Phys. Lett. A 223 (1996) 430.}
\item {[6]}{G.Casati, I.Guarneri, L.Molinari and K.\.Zyczkowski: "Periodic
Band Random matrices and conductance in disordered media", Phys. Rev. Lett. 72
(1994) 2697;\par
G.Casati, I.Guarneri and G.Maspero: "Landauer and Thouless 
conductance: a Band Random Matrix approach", J. Phys. I France 7 (1997) 729;
\par
S.Iida, A.Weidenm\"uller and J.A.Zuk: "Statistical scattering
theory, the supersymmetric method and universal conductance fluctuations",
Ann. of Phys. 200 (1990) 219.}
\item {[7]}{Ph. Jacquod and D.L.Shepelyansky: "Hidden Breit-Wigner 
distribution and other properties of random band matrices with preferential 
basis", Phys. Rev. Lett. 75 (1995) 3501.}
\item {[8]}{E.Br\'ezin and A.Zee: "Lattices of matrices", Nucl. Phys. B 441
[FS] (1995) 409;\par
S.Hikami and A.Zee: "Complex random matrix models with possible
applications to spin-impurity scattering in quantum Hall fluids", Nucl. Phys.
B 446 [FS] (1995) 337;\par
E.Br\'ezin, S.Hikami and A.Zee: "Oscillating density of states near
zero energy for matrices made of blocks with possible application to the random
flux problem", Nucl. Phys. B 464 [FS] (1996) 411.}
\item {[9]}{A.Crisanti, G.Paladin and A.Vulpiani: "Products of random 
matrices in statistical physics", Springer Series in Solid State Sciences
vol. 104, Springer-Verlag 1993.}
\item {[10]}{D.J.Thouless: "A relation between the density of states and the 
range of localization for one-dimensional random systems", J. Phys. C: Solid
State Phys. 5 (1972) 77.}
\item {[11]}{J.B.Pendry: "Symmetry and transport of waves in one-dimensional 
disordered systems", Advances in Physics 43 (1994) 461.}
\item {[12]}{M.Isopi and C,M.Newman: "The triangle law for Lyapunov exponents
of large random matrices", Comm. Math. Phys. 143 (1992) 591.}
\item {[13]}{J.Cook and B.Derrida: "Lyapunov exponents of large, sparse random
matrices and the problem of directed polymers with complex random weights",
J. Stat. Phys. 61 (1990) 961.}
\item {[14]}{C.W.Beenakker: "Random-matrix theory of quantum transport", 
Rev. Mod. Phys. 69 (1997) 731.}
\item {[15]}{A.MacKinnon and B. Kramer: "One parameter scaling of 
localization length and conductance in disordered systems", Phys. Rev. Lett.
47 (1981) 1546.}
\item {[16]}{P.Markos: "Universal scaling of Lyapunov exponents", J. Phys. 
A 30 (1997) 3441.}
\item {[17]}{T.Kottos, A.Politi, F.M.Izrailev and S.Ruffo: "Scaling
properties of Lyapunov spectra for the band random matrix model", Phys. Rev.
E 53 (1996) R5553;\par
T.Kottos, A.Politi and F.M.Izrailev: "Finite size corrections in Lyapunov 
spectra for Band Random Matrices", cond-mat/9801223.}
\item {[18]}{L.Molinari: "Transfer matrices and tridiagonal block 
Hamiltonians with periodic and scattering boundary conditions", J. Phys.
A: Math. Gen. 30 (1997) 983.}
\item {[19]}{N.Hatano and D.R.Nelson: "Localization transition in quantum 
mechanics", Phys. Rev. Lett. 77 (1996) 570.}
\item {[20]}{I.Goldsheid and B.Khoruzhenko: "Distribution of eigenvalues
in Non-Hermitian Anderson models", Phys. Rev. Lett. 80 (1998) 2897;\par
J.Feinberg and A.Zee: "Spectral curves of Non-Hermitean
Hamiltonians", cond-mat/9710040.}
\item {[21]}{V.Gurarie and A.Zee: "Localization length from single-particle
properties in disordered electronic systems", cond-mat/9802042.\par
N.M.Schnerb and D.R.Nelson: "Winding numbers, complex currents,
and Non-Hermitian localization", Phys. Rev. Lett. 80 (1998) 5172.}
\item {[22]}{P.W.Brouwer, P.G.Silvestrov and C.W.J.Beenakker: "Theory of
directed localization in one dimension", Phys. Rev. B 56 (1997) R4333.}
\item {[23]}{A.I.Aptekarev and E.M.Nikishin: "The scattering problem for a 
discrete Sturm-Liouville operator", Math. USSR Sbornik 49 (1984) 325.}
\item {[24]}{K.B.Efetov: "Quantum disordered systems with a direction",
Phys. Rev. B 56 (1997) 9630.}
\item {[25]}{J.Feinberg and A.Zee: "Non-hermitian random matrix theory: method
of hermitian reduction", Nucl. Phys. B 504 [FS] (1997) 579.}
\vfill\bye